\documentclass[final,5p,times,twocolumn]{elsarticle}
\usepackage{xcolor}
\usepackage{amssymb}
\usepackage{amsmath}
\usepackage{ulem}
\usepackage{scrextend}
\usepackage{threeparttable} 
\journal{Physics Letters B}

\begin{document}

\begin{frontmatter}

\title{
Shape Coexistence in $^{94}$Zr from a Model-Independent Analysis}

\author[a,b]{N.~Marchini\corref{cor1}}
\ead{naomi.marchini@fi.infn.it}
\author[a]{M.~Rocchini}
\author[c]{M.~Zieli\'nska}
\author[a]{A.~Nannini}
\author[d]{D.T.~Doherty}
\author[e]{N.~Gavrielov\corref{cor2}\fnref{fn0}}
\ead{noamgavrielov@gmail.com}
\author[f,g]{P.E.~Garrett}
\author[h]{K.~Hady\'nska-Kl\c{e}k}
\author[i]{A.~Goasduff}
\author[j]{D.~Testov\fnref{fn1}}
\author[d,j]{S.D.~Bakes}
\author[j,k]{D.~Bazzacco}
\author[l]{G.~Benzoni}
\author[d]{T.~Berry}
\author[i]{D.~Brugnara\fnref{fn2}}
\author[l,m]{F.~Camera}
\author[d]{W.N.~Catford}
\author[a]{M.~Chiari}
\author[j]{F.~Galtarossa}
\author[a]{N.~Gelli}
\author[i]{A.~Gottardo}
\author[i]{A.~Gozzelino}
\author[i]{A.~Illana\fnref{fn3}}
\author[n]{J.~Keatings}
\author[j,k]{D.~Mengoni}
\author[d]{L.~Morrison}
\author[i]{D.R.~Napoli}
\author[a]{M.~Ottanelli}
\author[a,b]{P.~Ottanelli}
\author[j,k]{G.~Pasqualato}
\author[j,k]{F.~Recchia}
\author[o,p]{S.~Riccetto\fnref{fn4}}
\author[n]{M.~Scheck}
\author[j,k]{M.~Siciliano\fnref{fn5}}
\author[i]{J.J.~Valiente~D\'obon\fnref{fn6}}
\author[j,k]{I.~Zanon\fnref{fn7}}

\affiliation[a]{organization={INFN Sezione di Firenze}, 
            city={Firenze},
            postcode={I-50019}, 
            country={Italy}}
\affiliation[b]{organization={Dipartimento di Fisica, Università di Firenze}, 
            city={Firenze},
            postcode={I-50019}, 
            country={Italy}}
\affiliation[c]{organization={IRFU, CEA, Université Paris-Saclay}, 
            city={Gif-sur-Yvette},
            postcode={F-91191}, 
            country={France}}
\affiliation[d]{organization={School of Mathematics and Physics, University of Surrey}, 
            city={Guildford}, 
            postcode={GU2 7XH}, 
            country={United Kingdom}}
\affiliation[e]{organization={GANIL, CEA/DRF-CNRS/IN2P3}, 
            city={Caen},
            postcode={F-14076},
            country={France}}
\affiliation[f]{organization={Department of Physics, University of Guelph},
            city={Guelph}, 
            postcode={N1G 2W1},
            country={Canada}}
\affiliation[g]{organization={Department of Physics, University of the Western Cape}, city={Bellville}, postcode={ZA-7535}, country={South Africa}}
\affiliation[h]{organization={Heavy Ion Laboratory, University of Warsaw},
            city={Warsaw},
            postcode={PL-02-093},
            country={Poland}}
\affiliation[i]{organization={INFN Laboratori Nazionali di Legnaro}, 
            city={Legnaro},
            postcode={I-35020}, 
            country={Italy}}
\affiliation[j]{organization={INFN Sezione di Padova}, 
            city={Padova},
            postcode={I-35131}, 
            country={Italy}}
\affiliation[k]{organization={Dipartimento di Fisica e Astronomia, Università degli Studi di Padova}, 
            city={Padova},
            postcode={I-35131}, 
            country={Italy}}
\affiliation[l]{organization={INFN Sezione di Milano}, 
            city={Milano},
            postcode={I-20133}, 
            country={Italy}}
\affiliation[m]{organization={Dipartimento di Fisica, Università di Milano}, 
            city={Milano},
            postcode={I-20133}, 
            country={Italy}}
\affiliation[n]{organization={Department of Physics, University of the West of Scotland},
            city={Paisley},
            postcode={G72 0LH},
            country={United Kingdom}}
\affiliation[o]{organization={Dipartimento di Fisica e Geologia, Università degli Studi di Perugia},
            city={Perugia},
            postcode={I-06123},
            country={Italy}}
\affiliation[p]{organization={INFN Sezione di Perugia},
            city={Perugia},
            postcode={I-06123},
            country={Italy}}

\cortext[cor1]{Corresponding author}
\cortext[cor2]{Corresponding author}

\fntext[fn0]{Present address: Department of Physics, Nuclear Research Center Negev, Be'er Sheva 84190, Israel}
\fntext[fn1]{Present address: Extreme Light Infrastructure-Nuclear Physics (ELI-NP), National Institute for Physics and Nuclear Engineering ``Horia Hulubei'', 077125 M\u{a}gurele-Bucharest, Romania}
\fntext[fn2]{Present address: Gesellschaft for Schwerionenforschung (GSI) mbH, 64291 Darmstadt, Germany}
\fntext[fn3]{Present address: Grupo de F\'isica Nuclear (GFN) and IPARCOS, Universidad Complutense de Madrid, CEI Moncloa, E-28040 Madrid, Spain}
\fntext[fn4]{Present address: Department of Physics, Queen’s University, ON K7L 3N6 Kingston, Canada}
\fntext[fn5]{Present address: Physics Division, Argonne National Laboratory, Lemont, 60439, IL, USA}
\fntext[fn6]{Present address: Instituto de Física Corpuscular (IFIC), E-46980 Paterna, España }
\fntext[fn7]{Present address: Department of Physics, Stockholm University, Stockholm, 10691, Sweden}

\begin{abstract}

Low-lying states of $^{94}$Zr were investigated via low-energy multi-step Coulomb excitation. From the measured $\gamma$-ray yields, \textcolor{black}{16} reduced   \textcolor{black}{E2} transition probabilities between low-spin states were determined, together with the spectroscopic quadrupole moments of the $2_{1,2}^+$ states. Based on this information, for the first time in the Zr isotopic chain, the shapes of the $0_{1,2}^+$ states including their deformation softness were inferred in a model-independent way using the quadrupole sum rules approach. The ground state of $^{94}$Zr possesses a rather diffuse shape associated with a spherical configuration, while the $0_2^+$ state is \textcolor{black}{triaxial tending towards} oblate and more strongly deformed. {\color{black}The observed features of shape coexistence in $^{94}$Zr are consistent with both Monte-Carlo shell-model predictions and IBM-CM calculations, and provide model-independent constraints on the shape character assigned in the IBM-CM to the intruder configuration in $^{92\text{--}96}$Zr.}

\end{abstract}

\begin{keyword}
Low-energy Coulomb excitation \sep Shape coexistence \sep Nuclear structure

\end{keyword}

\end{frontmatter}

\section{Introduction}

In the nuclear landscape characterized by gently evolving ground-state properties, the region of strontium and zirconium nuclei with neutron number $N$ close to 60 stands out. Nowhere else is the transition from a spherical to a strongly deformed ground state occurring as rapidly as when passing from $N=58$ $^{96}$Sr, $^{98}$Zr to $N=60$ $^{98}$Sr, $^{100}$Zr~\cite{campbell,garrett2022experimental,Leoni2024}, and a reproduction of this unique behaviour had been a long-standing challenge for nuclear-structure theory~\cite{federmanpittel,skalski1993,ozen2006,sieja2009,rodriguezguzman2010,mei2012,petrovici2012,xiang2012,nomura2016}. This dramatic shape change is often interpreted as a quantum phase transition (QPT)~\cite{cejnar2010quantum,fortunato2021}, in analogy to more familiar thermodynamic phase changes seen in many macroscopic systems. The transition from one shape phase to another corresponds to structural rearrangements in the nucleus, which are reflected in observables such as, e.g., two-nucleon separation energies, charge radii, level energies, and transition rates (see, e.g., Refs.~\cite{johansson,cheifetz1970experimental,thayer2003collinear,ansari2017, AME2020,giorgia2023}).

Recently, two theoretical approaches~\cite{togashi2016quantum,garciaramos2019,garciaramos2020,gavrielov2022zr} succeeded to describe the abrupt change in deformation observed at $N=58-60$ in the Zr isotopic chain, both of them invoking the QPT concept. The Monte-Carlo shell-model (MCSM) calculations~\cite{togashi2016quantum} attributed the shape transition to the so-called type-II shell evolution mechanism~\cite{otsuka2016,otsuka2018quantum}, being an example of self-organization of macroscopic systems observed in many domains of physics and beyond~\cite{synergetics}. This mechanism involves modifications of nuclear effective single-particle energies due to the occupation of specific orbitals. In the Zr nuclei, tensor and central forces act coherently to substantially lower the energies of proton-neutron spin-orbit partners, closing the pronounced spherical subshell gaps. The reorganized subshell sequence favors larger deformation thanks to coherent contributions of the configurations involved (Jahn-Teller effect~\cite{jahnteller}). According to the MCSM calculations, the deformed configurations assume a variety of shapes throughout the Zr isotopic chain that include prolate, oblate and triaxial ones~\cite{togashi2016quantum}.

A large set of experimental spectroscopic data related to the shape transition in the Zr isotopes was also satisfactorily reproduced in the framework of configuration mixing within the interacting boson model (IBM-CM)~\cite{garciaramos2019,garciaramos2020,gavrielov2022zr}. With the IBM-CM, another QPT of shape-evolution was identified in the Zr intruder configuration, on top of the QPT in the configuration content of the ground state also suggested by MCSM~\cite{togashi2016quantum}. This evolution in structure was dubbed as \textit{intertwined} QPT (IQPT)~\cite{gavrielov2022zr}. In an IQPT, two configurations are present, each of which may undergo an individual QPT. For $N<60$, the IBM-CM calculations of Ref.~\cite{gavrielov2022zr} predict a spherical vibrational (possessing good U(5) symmetry) configuration for the ground state, coexisting with a weakly deformed (quasi-U(5) symmetry) intruder configuration. The two configurations interchange at $N=60$, and in parallel the intruder configuration undergoes a QPT from weakly deformed to well deformed (good rotational SU(3) symmetry). Similar to the MCSM calculations, the IBM-CM reproduces the observed small mixing~\cite{kremer2016first} between the normal and intruder configurations \textcolor{black}{for the $0^+$ and $2^+$ states}.

While the experimental information on the nature of $0_2^+$ states in the Zr isotopes with $N\geq60$~\cite{mach1989,urban2019,wu2024,hill1991} is limited due to the neutron-rich character and short half-lives of the nuclei in question, definitive conclusions regarding their character in $N<60$ Zr nuclei are not possible either. For example, the results of lifetime measurements for excited states in $^{98}$Zr~\cite{singh98Zr} were interpreted, with a guidance from MCSM calculations, in terms of triple shape coexistence with notably a moderately deformed $0^+_2$ state and a rotational structure built on it. At the same time, a different set of lifetimes~\cite{karayonchev2020tests} measured in the same nucleus was found to be in general agreement with a vibrational interpretation of states built on the $0_2^+$ state. Both theoretical approaches, however, do not reproduce the large $B(E2;4_1^+ \to 2_2^+)$ value deduced from the data~\cite{singh98Zr,karayonchev2020tests}. Moreover, in $^{94,96}$Zr enhanced $B(E2;2_2^+ \to 0_2^+)$ values of 19(2)~W.u.~\cite{chakraborty2013collective} and 36(11)~W.u.~\cite{kremer2016first}, respectively, were determined and interpreted as corresponding to in-band transitions in collective rotational structures built on the deformed $0_2^+$ states, thus supporting a shape-coexistence scenario.  \textcolor{black}{A very recent study \cite{Zielinska_PLB} of $^{96}$Zr combining Coulomb excitation and $\beta$-decay measurements resulted in a refined $B(E2;2^+_2\rightarrow 0^+_2)$ value of 38.9(57) W.u.  With this more precise $B(E2)$ value, combined with the observations from transfer reactions, it was argued that the $0^+_2$ state in $^{96}$Zr has a different configuration than the $0^+_2$ state in $^{94}$Zr, supporting a triple shape-coexistence scenario in the Zr isotopes.}  However, an 
evaluation~\cite{witt2019data} of the 
experimental data for $^{96}$Zr showed that a vibrational character of the structure built on the $0_2^+$ state cannot be excluded. We note that the moderate values of the $E0$ transition strength, quantified by $10^3\times\rho^2(E0)$, observed for $N\leq 58$ can be consistent with both a shape-coexistence scenario and with transitions between vibrational states involving a change in phonon (or $d$-boson) number of 0 or $\pm 2$~\cite{WoodE0}.

The above examples show the limitations of the current experimental knowledge of the even-even Zr nuclei and the difficulties in distinguishing between various interpretations. In particular, the MCSM results tend to be presented in the context of the underlying shapes, which can only be deduced from the existing experimental data using strong model assumptions.  Critical data, such as spectroscopic quadrupole moments, $Q_s$, of $2^+$ states that could discriminate between a spherical-vibrational and deformed-rotational interpretation, have been lacking. 

The present Letter reports not only the determination of $Q_s$ for the $2^+_1$ and $2^+_2$ states, but further provides the quantities $\langle Q^2\rangle$ and $\langle Q^3\cos3\delta\rangle$, which are representative of the nuclear shape, determined for the $0^+_1$ and $0^+_2$ states in $^{94}$Zr in a multi-step Coulomb-excitation study. These experimental results \textcolor{black}{, obtained without invoking any nuclear model,} establish coexistence of a quasi-spherical ground state and a more deformed \textcolor{black}{triaxial-}oblate structure built on the $0_2^+$ state\textcolor{black}{. They} are compared with new IBM-CM calculations building on those reported in Ref.~\cite{gavrielov2022zr}, as well as with the MCSM calculations~\cite{togashi2016quantum,Leoni2024}.

\section{Experiment}

The Coulomb-excitation technique is ideally suited to study collectivity of yrast and non-yrast states at low spin and excitation energy~\cite{magdaeuroschool}. The $^{94}$Zr nucleus was chosen among the Zr isotopes since it has a favorable level scheme with the states of interest below 2.5~MeV excitation energy. Moreover, the relatively large $B(E2;0_2^+ \to 2_1^+)=9.3(4)$~W.u.~value~\cite{chakraborty2013collective} results in an enhanced cross section to Coulomb excite the $0_2^+$ state via a two-step process, facilitating population of higher-lying states. Furthermore, high-intensity beams of the stable $^{94}$Zr can be produced, leading to data sets with high statistical quality.

\begin{figure}
   \centering
   \includegraphics[width=\columnwidth]{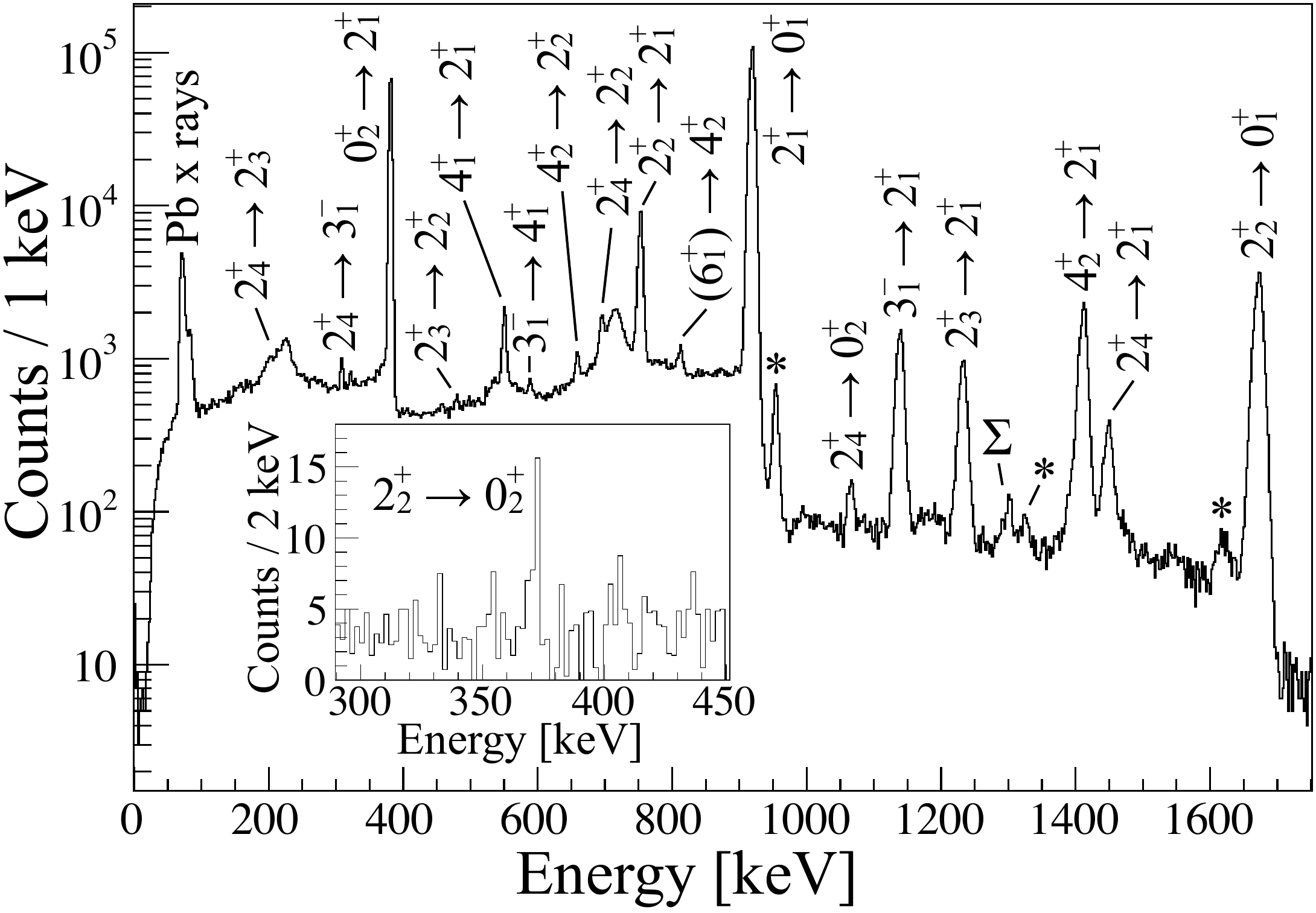}
   \caption{Portion of the $\gamma$-ray energy spectrum observed following Coulomb excitation of the $^{94}$Zr beam impinging on a $^{208}$Pb target, Doppler corrected for the projectile. Transitions in $^{95}$Zr are marked with asterisks and $\Sigma$ denotes a sum peak. The inset shows the $2_2^+ \to 0_2^+$ transition observed in $\gamma$-$\gamma$-particle coincidences (gated on the $0_2^+ \to 2_1^+$ $\gamma$-ray transition).}
   \label{fig:spectrum}
\end{figure}

\begin{figure}
   \centering
   \includegraphics[width=\columnwidth]{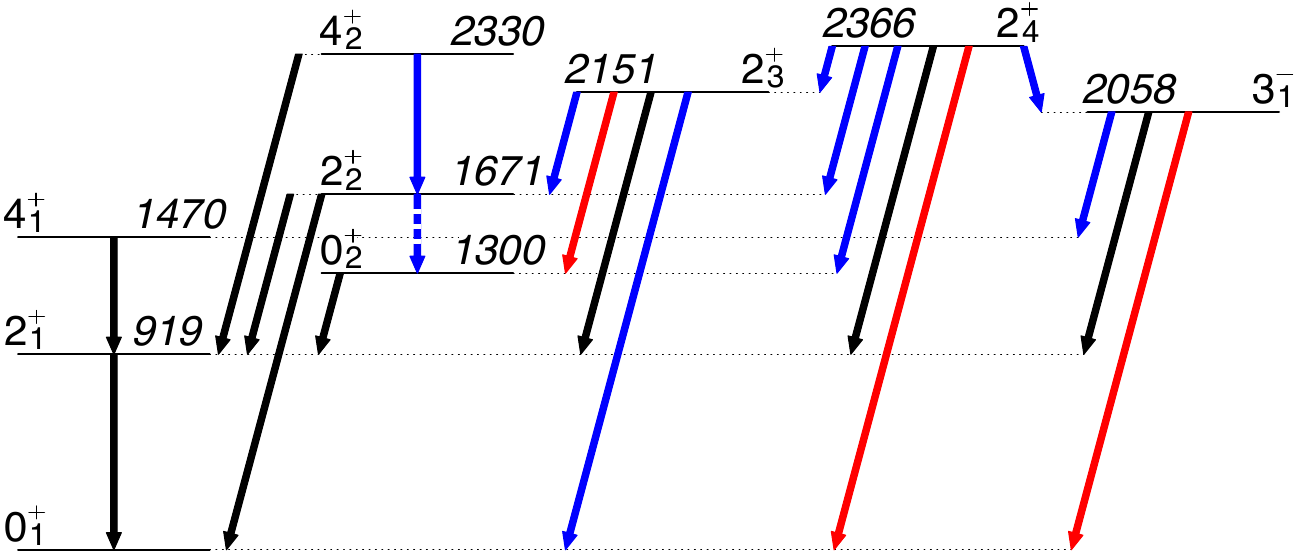}
   \caption{Partial level scheme of $^{94}$Zr showing transitions relevant for the present work. The level energies are given in keV. The number of counts observed for the transitions marked in black was sufficient to subdivide the data into eight angular ranges. For the transitions marked in blue, the total number of counts in coincidence with the SPIDER array was used in the analysis.  The $2_2^+ \to 0_2^+$ transition was observed only in $\gamma$-$\gamma$-particle coincidences (dashed blue). Transitions marked in red have not been observed in the present experiment, but the corresponding matrix elements were considered in the data analysis with the GOSIA code.}
   \label{fig:level_scheme}
\end{figure}

The experiment was performed at the INFN Legnaro National Laboratories (LNL). During 4 days of data taking, a 0.1-pnA beam of $^{94}$Zr impinged on a 0.97(2)-mg/cm$^2$ thick, self-supporting $^{208}$Pb target (the target thickness was measured at the INFN LABEC laboratory using the Rutherford back-scattering technique~\cite{rocchini2021applications}). The beam energy, 370~MeV, fulfilled Cline's ``safe''-energy criterion~\cite{cline1986nuclear} for the range of scattering angles observed, ensuring that the contribution from the nuclear forces to the excitation process was negligible. The $\gamma$ rays deexciting the populated states were detected by the GALILEO $\gamma$-ray spectrometer~\cite{goasduff2021galileo} combined with six large-volume $3'' \times 3''$ LaBr$_3$:Ce detectors~\cite{giaz2013characterization}, and the scattered $^{94}$Zr ions by the SPIDER silicon array~\cite{rocchini2020spider}. The absolute photo-peak efficiencies at 1332.5~keV of GALILEO and the LaBr$_3$:Ce array were 2.1\% and 1.1\%, respectively. SPIDER comprised seven silicon detectors with trapezoidal shapes, segmented eight-fold in polar angle and covering $\theta_{LAB}$ angles from 123$^\circ$ to 161$^\circ$. The standard GALILEO and SPIDER presorting and data-correction procedures~\cite{goasduff2021galileo,rocchini2020spider,rocchini2021onset} were implemented in the analysis, resulting in 8.7~keV full width at half maximum at 919~keV energy in the final Doppler-corrected $\gamma$-ray spectrum from GALILEO in coincidence with the back-scattered $^{94}$Zr ions. This spectrum is shown in Fig.~\ref{fig:spectrum}, while Fig.~\ref{fig:level_scheme} reports a partial $^{94}$Zr level scheme indicating the transitions observed in the present work. The inset of Fig.~\ref{fig:spectrum} displays the $2_2^+ \to 0_2^+$ transition that was only observed in the $\gamma$-$\gamma$-particle coincidence analysis combining the HPGe and LaBr$_3$:Ce data, confirming its placement proposed in Ref.~\cite{chakraborty2013collective}.
 For the unobserved  $2_4^+ \to 0_1^+$ transition, for which the branching ratio is unknown, observation upper limit were imposed following the method outlined in Ref. \cite{currie}. 
In addition to transitions in $^{94}$Zr, a few weaker lines in the spectrum, at the level of 0.3\% or less of the intensity of the $2^+_1\to 0^+_1$ $^{94}$Zr transition, arise from sub-barrier one-neutron transfer to $^{95}$Zr. The $^{95}$Zr levels identified correspond to those with the largest spectroscopic factors for one-neutron transfer~\cite{Bingham}, and the observed populations are in line with the transfer cross sections measured in, e.g., Ref.~\cite{kernan1991} for conditions in which the nuclear interaction had a negligible influence on the excitation process.

\section{Coulomb-excitation analysis}

The  yields in $^{94}$Zr \textcolor{black}{measured with GALILEO} were divided into the eight angular ranges defined by the SPIDER segmentation~\cite{rocchini2020spider} to exploit the angular dependence of the Coulomb-excitation cross sections, except for weaker transitions for which the coincidences with the entire SPIDER array were considered (see Fig.~\ref{fig:level_scheme}). 

\begin{figure}
   \includegraphics[width=\columnwidth]{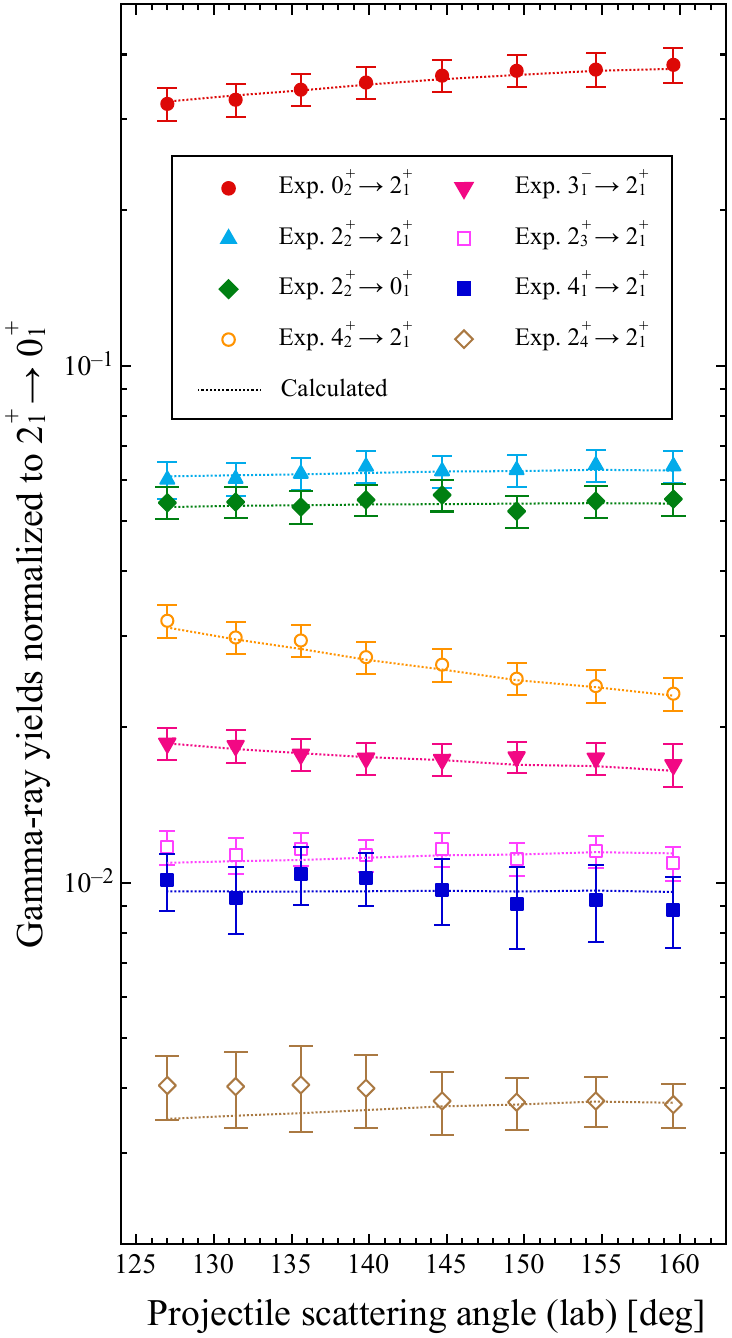}\label{fig:gamma-yields}
   \caption{Comparison between the $\gamma$-ray yields measured in the present experiment for the eight projectile angular ranges defined by the SPIDER geometry and those calculated using the final set of reduced electromagnetic matrix elements resulting from the $\chi^2$ minimization performed with the GOSIA code, listed in Table~\ref{tab:transition_prob}. The calculated yields are integrated over the target thickness and the particle detector angular coverage. All yields are efficiency corrected and given relative to those of the $2_1^+ \to 0_1^+$ transition. }
\end{figure}

\begin{figure}
   \includegraphics[width=\columnwidth]{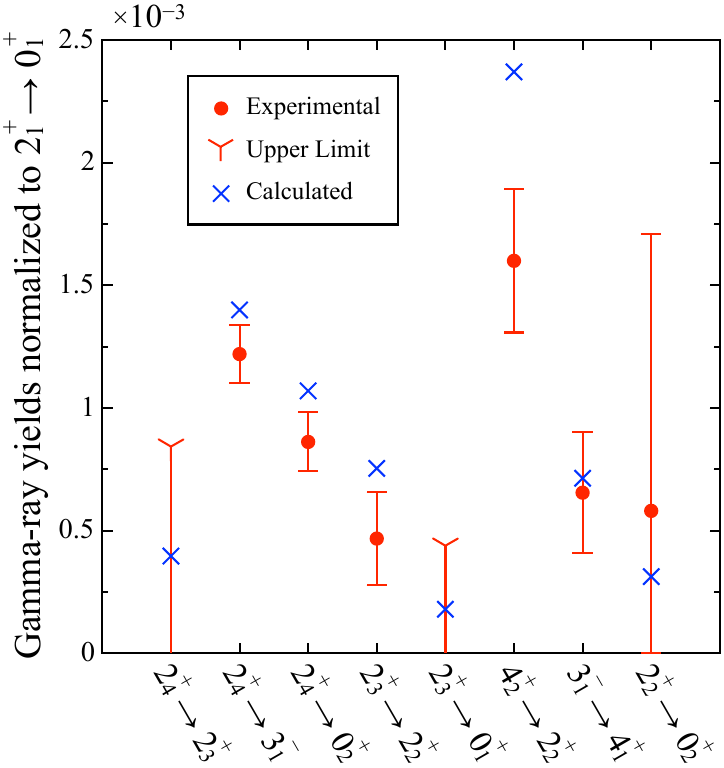}
   \caption{\label{fig:gamma-yields_tot} Comparison between the $\gamma$-ray yields measured in the present experiment analyzed in coincidence with the full SPIDER detector and those calculated using the final set of reduced electromagnetic matrix elements resulting from the $\chi^2$ minimization performed with the GOSIA code, listed in Table~\ref{tab:transition_prob}. The calculated yields are integrated over the target thickness and the particle detector angular coverage. All yields are efficiency corrected and given relative to those of the $2_1^+ \to 0_1^+$ transition. Upper limits on the observed intensities are given at the 95\% confidence level. }  
\end{figure}

The extracted numbers of $\gamma$-particle coincidences for the observed transitions, presented in Figs.~\ref{fig:gamma-yields},~\ref{fig:gamma-yields_tot}, were analyzed using the GOSIA code~\cite{czosnyka1983gosia}. \textcolor{black}{The $2^+_4\to2^+_3$,  $2^+_3\to0^+_1$, and $2^+_4\to2^+_2$ peaks are visible in the spectrum, but their areas cannot be fitted accurately. For the former two, their upper limits (at the 95\% confidence level) were determined, and the latter, positioned on the Compton edge of the intense 2$^+_1 \to 0^+_1$ transition, was excluded from the analysis.}

Figure~\ref{fig:gamma-yields} shows a comparison between the measured $\gamma$-ray yields (efficiency-corrected and normalized to the $2_1^+ \to 0_1^+$ yield), measured in coincidence with the scattered projectile, and those calculated from the obtained set of matrix elements, for the eight angular ranges defined by the SPIDER geometry. An agreement within the $\pm1\sigma$ uncertainty on the data points is obtained for the whole angular range. 
Figure~\ref{fig:gamma-yields_tot} shows the same comparison for the transitions analyzed in coincidence with the full SPIDER detector (also in this case efficiency-corrected and normalized to the $2_1^+ \to 0_1^+$ yield). 
\textcolor{black}{The agreement is, on average, about $1.2\sigma$, with the largest discrepancy (2.6$\sigma$) observed for the $4_2^+ \to 2_2^+$ transition.}

While the present experiment was insensitive to $B(M1)$ and $B(E1)$ values, the corresponding matrix elements were included in the multidimensional GOSIA fit together with those for $E2$ and $E3$ multipolarities, and their values were constrained by complementary spectroscopic data. The level lifetimes were taken from Refs.~\cite{chakraborty2013collective,peters2013level}, the $E2/M1$ mixing ratios from Ref.~\cite{elhami2008experimental}, and the branching ratios from Refs.~\cite{chakraborty2013collective,elhami2008experimental,fotiades2002high,singh1976study,mandal2020}. The final set of matrix elements obtained in the analysis reproduces all these values within $\pm 1\sigma$ uncertainty\textcolor{black}{, with an exception of the 2$^+_4$ lifetime, for which the current data point to a considerably lower value (2$\sigma$ disagreement)}.  
The intensity of the $(6^+_1) \to 4^+_2$ transition has been included in the analysis, but due to ambiguities~\cite{elhami2008experimental,pantelica2005,fotiades2002high} regarding the spin of the presumed (6$^+_1$) state at 3142~keV, the corresponding matrix element is not reported. \textcolor{black}{Influence of unknown quadrupole moments on the excitation cross sections, as well as the effects of possible excitations of unobserved states, were evaluated as described in Supplemental Material \cite{supplement} and included in the quoted uncertainties of deduced matrix elements.} Absolute normalization of the measured cross sections was achieved via inclusion in the fit of a weighted average of the 2$^+_1$ lifetimes~\cite{raman2001transition,horen1993half,lund1995isospin}\textcolor{black}{, equal to $10.0(4)$~ps,} and consequently the analysis was not sensitive to the $B(E2; 2_1^+ \to 0_1^+)$ value~\cite{magdaEPJA,magdaeuroschool}. The following sign convention was imposed for $E2$ matrix elements: those for in-band transitions were assumed to be positive, as well as those of $\langle 0^+_2 \| E2 \| 2^+_1 \rangle$, $\langle 2^+_1 \| E2 \| 2^+_3 \rangle$ and $\langle 2^+_1 \| E2 \| 2^+_4 \rangle$, with the signs of remaining matrix elements determined relative to them.
 
The present work determined $Q_s$ of the $2_{1,2}^+$ states in $^{94}$Zr (Table~\ref{tab:quadrupole_moments}) -- the first such determination in the Zr isotopic chain~\cite{Stone_quadtables}. Moreover, a large number of $E2$ matrix elements were precisely measured that are reported in Table~\ref{tab:transition_prob} together with the corresponding $B(E2)$ values. In addition, the $B(E3; 3_1^- \to 0_1^+)=\textcolor{black}{37(5)}$~W.u.~value was extracted from the data, which agrees with the evaluated value \textcolor{black}{$B(E3; 3_1^- \to 0_1^+)=24(8)$~W.u.~\cite{kibedi2002reduced} within 1$\sigma$}.

\begin{table}
\centering
   \caption{Diagonal $E2$ matrix elements measured in the present work and corresponding spectroscopic quadrupole moments, compared with those resulting from IBM-CM calculations.}
   {\renewcommand{\arraystretch}{1.5}
   \begin{tabular}{cccc}\\
      \hline
      $J_i$ &$\langle J_i \| E2 \| J_i \rangle$~[$e$b] &\multicolumn{2}{c}{$Q_s(J_i)$~[$e$b]} \\
      \cline{2-2}\cline{3-4}
      &Exp. & Exp. & IBM-CM \\
      \hline
$2_{1}^+$ & \textcolor{black}{$+0.17(3)$} &\textcolor{black}{$+0.13(2)$} & $+0.066$ \\
 $2_{2}^+$ & $\textcolor{black}{+0.32(5)}$ &$\textcolor{black}{+0.24(4)}$ & $+0.31$\\
      \hline
   \end{tabular}
   }
   \label{tab:quadrupole_moments}
\end{table}

\begin{table}[!htb]
\begin{threeparttable}
   \caption{\label{tab:transition_prob} Transitional $E2$ matrix elements obtained in the present work (with the exception of $\langle 0^+_1 \| E2 \| 2^+_1 \rangle$ provided for completeness) and corresponding $B(E2)$ values compared with those resulting from IBM-CM calculations.}
   {\renewcommand{\arraystretch}{1.5}
   \begin{tabular}{cccc}
      \hline
      $J_i \rightarrow J_f$ & $\langle J_f \| E2 \| J_i \rangle$~[$e$b] & \multicolumn{2}{c}{$B(E2; J_i \rightarrow J_f)$~[W.u.]} \\
      \cline{2-2}\cline{3-4}
      & Exp. & Exp. & IBM-CM \\ 
      \hline
      $2^+_1 \longrightarrow 0^+_1$ & $+0.250(7)$\tnote{a,b} & 4.8(2)\tnote{b} & 2.7 \\
      $0^+_2 \longrightarrow 2^+_1$ & $+0.155(4)$\tnote{a} & 9.5(5) & 9.3 \\
      $4^+_1 \longrightarrow 2^+_1$ & $+0.141(4)$\tnote{a} & 0.87(5) & ---\tnote{c} \\
      $2^+_2 \longrightarrow 0^+_2$ & \textcolor{black}{$+0.488(12)$}\tnote{a} & \textcolor{black}{18.8(9)} & 20.2 \\
      $2^+_2 \longrightarrow 2^+_1$ & \textcolor{black}{$+0.029^{+0.018}_{-0.017}$} & \textcolor{black}{$< 0.1$} & 1.49 \\
      $2^+_2 \longrightarrow 0^+_1$ & $\textcolor{black}{+0.222(6)}$ & \textcolor{black}{3.9(2)} & 0.82 \\
      $4^+_2 \longrightarrow 2^+_2$ & $\textcolor{black}{+1.02(3)}$\tnote{a} & \textcolor{black}{45(3)} & 26.6 \\
$4^+_2 \longrightarrow 2^+_1$ & $\textcolor{black}{+0.65(2)}$ & \textcolor{black}{18.4(11)}  & 2.1 \\
      $2^+_3 \longrightarrow 2^+_2$ & $\textcolor{black}{+0.32(9)}$ & $\textcolor{black}{8(5)}$ & 17.3 \\
      $2^+_3 \longrightarrow 0^+_2$ & $\textcolor{black}{\pm0.11^{+0.04}_{-0.05}}$
       & $\textcolor{black}{1.0^{+0.8}_{-0.7}}$ 
      & 0.07 \\
    $2^+_3 \longrightarrow 2^+_1$ & $\textcolor{black}{+0.28(4)}$\tnote{a} & $\textcolor{black}{6.2(18)}$ & 1.2 \\ 
      $2^+_3 \longrightarrow 0^+_1$ & $\pm0.018\textcolor{black}{(3)}$ & $0.026\textcolor{black}{(9)}$ & 0.001 \\
 $2^+_4 \longrightarrow 2^+_3$ & $\textcolor{black}{<0.16}$ & $\textcolor{black}{<2}$ & 2.44 \\
      $2^+_4 \longrightarrow 2^+_2$ & $\pm \textcolor{black}{0.11(5)}$ & $\textcolor{black}{{1.0^{+1.0}_{-0.7}} }$ & 0.1 \\
      $2^+_4 \longrightarrow 0^+_2$ & \textcolor{black}{$\pm 0.32(2)$} & $\textcolor{black}{8.1(10)}$ & 0.06 \\
      $2^+_4 \longrightarrow 2^+_1$ & $\textcolor{black}{+0.17(3)}$\tnote{a} & $\textcolor{black}{2.3(8)}$ & 0.001 \\
          $2^+_4 \longrightarrow 0^+_1$ & $\textcolor{black}{-0.012(9)}$ & $\textcolor{black}{<4\cdot10^{-2}}$ & $3\cdot10^{-4}$ \\
      \hline
   \end{tabular}
   }
   \begin{tablenotes}
   \footnotesize
   \item[a] Sign imposed in the analysis (see text for details).
   \item[b] Determined from literature data.
   \item[c] Outside IBM-CM model space (see Ref.~\cite{gavrielov2022zr}).
\end{tablenotes}
\end{threeparttable}
\end{table}

\section{Experimental quadrupole shape invariants}

The different deformations of two structures in $^{94}$Zr are evidenced by the large difference between the spectroscopic quadrupole moments measured for the $2^+_{1,2}$ states (Table~\ref{tab:quadrupole_moments}). To get more precise, model-independent information on the nuclear charge distribution in the ground state and the first excited $0^+$ state, a quadrupole sum rule analysis~\cite{cline1986nuclear,kumar1972intrinsic,magdaeuroschool} was applied to the obtained set of $E2$ matrix elements. The electric quadrupole operator, ${\mathcal{M}}(E2)$, can be expressed in the principal-axis frame using two parameters:
\begin{eqnarray} 
{\mathcal{M}}(E2,{\mu=0})=&Q\cos\delta \nonumber \\
{\mathcal{M}}(E2,{\mu=\pm2})=&\displaystyle\frac{1}{\sqrt{2}}Q\sin\delta.
\label{eq:Q-delta}
\end{eqnarray}
By definition, in this frame of reference the ${\mathcal{M}}(E2,\mu=\pm1)$ components vanish. The parameters $Q$ and $\delta$ are analogous to the deformation parameters 
$\beta_2$ and $\gamma$, but instead of the mass distribution they represent the quadrupole charge distribution. 
The products of the $E2$ operators coupled to zero angular momentum are rotationally invariant and thus their expectation values can be expressed by the $Q$ and $\delta$ on one hand, and by products of  $\langle I_j \| {\mathcal{M}}(E2)\| I_i \rangle$ $E2$ matrix 
elements on the other hand. For the lowest-order invariant:
\begin{equation}
    \left\{ {\mathcal{M}}(E2) \times {\mathcal{M}}(E2)\right\}^0=\frac{1}{\sqrt{5}}Q^2
\end{equation}
its expectation value for a state $J_n$, related to the overall deformation, can be expressed through $E2$ matrix elements $M_{ab}=\langle J_a || E2 || J_b \rangle$ as follows:
\begin{equation}
    \langle J_n | Q^2 | J_n \rangle=\frac{\sqrt{5}\left( -1 \right)^{2J_n}}{\sqrt{2J_n+1}}\sum _i  M_{ni}M_{in} \begin{Bmatrix} 2 & 2 & 0 \\ J_n & J_n & J_i \end{Bmatrix}
\end{equation}
The second-order invariant can be expressed as:
\begin{equation}
    \left\{ \left[{\mathcal{M}}(E2) \times {\mathcal{M}}(E2)\right]^2\times {\mathcal{M}}(E2) \right\}^0=-\sqrt{\frac{2}{35}}Q^3\cos 3\delta,
\end{equation}
while the evaluation using the intermediate-state expansion yields:
\begin{multline}
    \langle J_n | Q^3 \cos 3\delta | J_n \rangle = \\ - \sqrt{\frac{35}{2}}\frac{(-1)^{2J_n}}{2J_n+1}\sum _{ij} M_{ni}M_{ij}M_{jn} \begin{Bmatrix} 2 & 2 & 2 \\ J_n & J_j & J_i \end{Bmatrix}
\end{multline}
The triaxial parameter $\delta$ giving the quadrupole asymmetry, \textit{i.e.}, the deviation from axial symmetry, was derived under the assumption:
\begin{equation}
    \langle Q^3 \cos 3\delta \rangle \cong \langle Q^2 \rangle^{3/2}\langle \cos 3\delta \rangle
\end{equation}
Specifically, $\langle \cos \left( 3\delta \right) \rangle$ is $+1$ for a prolate shape ($\langle \delta \rangle=0^\circ$), 0 for a maximally triaxial shape ($\langle \delta \rangle=30^\circ$), and $-1$ for an oblate shape ($\langle \delta \rangle=60^\circ$).

The third-order invariant can be formed with different intermediate $J$ couplings, namely $J=0,2,4$, which involve summation over different sets of reduced $E2$ matrix elements:

\begin{equation}
    \left\{ \left[{\mathcal{M}}(E2) \times {\mathcal{M}}(E2)\right]^J \times \left[ {\mathcal{M}}(E2) \times {\mathcal{M}}(E2) \right] ^J \right\} =\frac{1}{5}Q^4 \label{eq:q4}
\end{equation}

These three couplings should yield the same $\langle Q^4 \rangle$ values, providing a consistency test for the set of $E2$ matrix elements, as well as for the convergence of the rotational-invariant sum rules themselves. 
\textcolor{black}{However, each of these three $\langle Q^4 \rangle$ values is evaluated independently from the remaining two, on the basis of a different subset of $E2$ matrix elements.   Hence, 
even though only the $Q^4(J=0)$ value could be determined from the present data set, there is no possible bias coming from missing couplings, as all relevant matrix elements (or their limits) involving states up to 2$^+_4$ were known. The following expression was used:}
\begin{multline}
    \langle J_n | Q^4 | J_n \rangle = \\ \frac{\textcolor{black}{5}}{2J_n+1}\sum _{ijk} M_{ni}M_{ij}M_{jk}M_{kn} \begin{Bmatrix} 2 & 2 & 0 \\J_n & J_j & J_i \end{Bmatrix} \\ \begin{Bmatrix} 2 & 2 & 0 \\ J_n & J_j & J_k \end{Bmatrix} (-1)^{J_n-J_j}
\end{multline}

The expectation values of $Q^2$ and $Q^4$ were then used to construct the quantity $\sigma({Q^2})=\sqrt{\langle Q^4 \rangle-(\langle Q^2 \rangle)^2}$, i.e. the softness of the overall deformation given by $Q^2$.

To determine the quadrupole invariants for a 0$^+$ state, it is necessary to know the $E2$ matrix elements connecting the 0$^+$ state of interest to all relevant 2$^+$ states for the first-order invariant. Additionally, for the second-order invariant, the quadrupole moments of the 2$^+$ states with their absolute signs, as well as the relative signs of the transition matrix elements, are required; \textcolor{black}{the latter also affect the $Q^4(J=0)$. In the present analysis, contributions from products of $E2$ matrix elements involving those with unknown relative signs were included in the uncertainties of the determined invariants. Regarding the unknown quadrupole moments of the 2$^+_{3,4}$ states, a worst-case scenario was considered in which their absolute values were equal to the quadrupole moment of the deformed 2$^+_{2}$ state plus 1$\sigma$, and the resulting contributions were included in the $\langle \cos 3 \delta \rangle$ uncertainties.} 

\begin{figure}
   \centering
   \includegraphics[width=0.8\columnwidth]{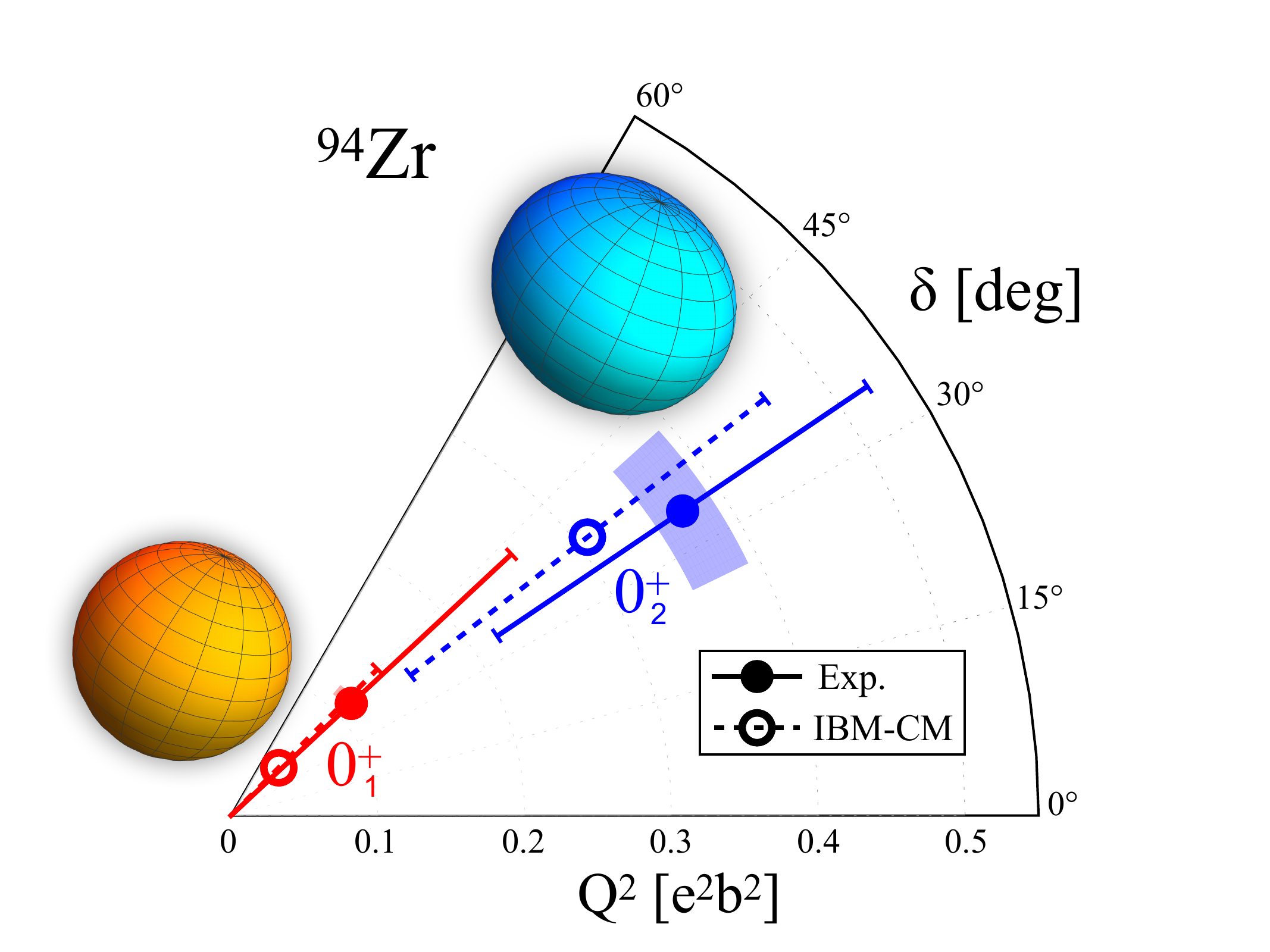}
   \caption{Shape parameters for the $0_{1,2}^+$ states in the ($Q^2$, $\delta$) plane resulting from the sum-rules analysis. The experimental results are shown with full circles, while the shaded areas represent their uncertainties (for the $0_1^+$ state the respective area is smaller than the marker size). 
   The results of the IBM-CM calculations are shown with empty circles. The lines (solid for the experimental values and dashed for the IBM-CM calculations) represent $\sigma({Q^2})$ dispersions. The results are reported in red for the $0_1^+$ state and in blue for the $0_2^+$ state. The illustrative shapes are plotted using $\beta_2$ and $\gamma$ values derived from the experimental invariant quantities for the $0_2^+$ state (blue ellipsoid), and $\beta_2=0$ for the $0_1^+$ state (orange sphere), see text for details.}
   \label{fig:shapes}
\end{figure}

The results of this approach are presented in Table~\ref{tab:sum_rules} and Fig.~\ref{fig:shapes}, with contributions of specific products of matrix elements to the obtained invariant quantities listed in the Supplemental Material \cite{supplement}. \textcolor{black}{The statistical uncertainties correspond to about 30\% of the quoted values, while the remaining part results from an evaluation of the influence of excitation of higher-lying states and unknown values of the quadrupole moments of the 2$^+_{3,4}$ states, and, for the $\langle Q^3 \cos \left( 3\delta \right) \rangle$ and $\langle Q^4 \rangle$ invariants, from the unknown relative signs of transitional matrix elements.} 

The $\langle Q^2 \rangle$ of the $0_2^+$ state is $\approx \textcolor{black}{3.3}$ times larger than that of the $0_1^+$ state. However, the fact that the dispersion of $\langle Q^2 \rangle$ for the $0_1^+$ state is comparable with the $\langle Q^2 \rangle$ value itself indicates that the ground state of $^{94}$Zr exhibits fluctuations about a spherical shape. In contrast, $\sigma({Q^2})$ for the $0_2^+$ state is considerably smaller than $\langle Q^2 \rangle$, which suggests a moderate static deformation. More insight into the shape of the $0_2^+$ state can be obtained from its value of $\langle \cos \left( 3\delta \right) \rangle$ \textcolor{black}{(equivalent to $\langle \delta \rangle={34}(8)^{\circ}$), which indicates a triaxial deformation tending towards an oblate shape, as indicated by} the positive value of the $2^+_2$ spectroscopic quadrupole moment. For the $0_1^+$ state, the $\langle \cos \left( 3\delta \right) \rangle$ value is of limited relevance since $\sigma({Q^2}) > \langle Q^2\rangle$.

\begin{table}[!hbt]
   \caption{Results of the quadrupole sum-rules analysis of experimental $E2$ matrix elements obtained in the present work (Exp.) compared with IBM-CM calculations (Th.).}
      {\renewcommand{\arraystretch}{1.5}
      \setlength{\tabcolsep}{4pt}
   \begin{tabular}{ccccccc}
      \hline
      $J_i$ & \multicolumn{2}{c}{$\langle{Q^2}\rangle$ [$e^2\text{b}^2$]} & \multicolumn{2}{c}{$\sigma({Q^2})$ [$e^2\text{b}^2$]} & \multicolumn{2}{c}{$\langle{\cos(3\delta)}\rangle$} \\
      \cline{2-3}\cline{4-5}\cline{6-7}
      & Exp. & Th. & Exp. & Th. & Exp. & Th. \\
      \hline
      $0^+_1$ & $\textcolor{black}{0.112(4)}$ & $0.046$ & $\textcolor{black}{0.147(11)}$ & $0.094$ & $\textcolor{black}{-0.64(22)}$ & $-0.72$\\
      $0^+_2$ & \textcolor{black}{0.371(21)} & $0.308$ & $\textcolor{black}{0.15(6)
      }$ & $0.153$ & $\textcolor{black}{-0.2(4)}$ & $-0.4$\\
      \hline
   \end{tabular}
   }
   \label{tab:sum_rules} 
\end{table}

\section{IBM-CM calculations}

The experimental results were compared with IBM-CM calculations building on Ref.~\cite{gavrielov2022zr}. \textcolor{black}{Using the $Q_s(2_{1,2}^+)$ values measured in the present 
work it, was possible to fix the $\chi$ model parameter to 
$\chi=+0.6$ instead of the previously adopted 
$\chi=-0.6$, without modifying any other parameter from 
Ref.~[24].} This choice does not change the spectrum reported in Ref.~\cite{gavrielov2022zr} but is responsible for the prolate-oblate nature of the deformation. 
\textcolor{black}{The new adopted sign for $\chi$ has also 
affected the sign of other members of the isotopic chain, 
$^\text{92--96}$Zr, for the calculation of the spectrum of 
the odd-mass $^\text{93--103}$Zr \cite{noam2025}.}
The calculated $B(E2)$ values are shown in Table~\textcolor{black}{\ref{tab:transition_prob}}. 
\textcolor{black}{A 
simple measure of the agreement is provided by the ratio
$R = \frac{B(E2)_{\rm IBM\!-\!CM}}{B(E2)_{\rm exp}}.$
For the main collective transitions $0^+_2 \rightarrow 
2^+_1$, $2^+_2 \rightarrow 0^+_2$ we obtain $R = 0.98, 1.07,$
respectively, i.e., a deviation of $\lesssim 7\%$. The 
ground-state transition $2^+_1 \rightarrow 0^+_1$ is 
reproduced within a factor of $\simeq 2$ ($R = 
0.56$), which reflects the fact that the effective $E2$ 
boson charges are constant along the Zr chain rather than 
adjusted to $^{94}$Zr alone. Similar level of reproduction ($R \simeq 0.59$) is obtained for the $4^+_2 
\rightarrow 2^+_2$ transition, while $4^+_2 \rightarrow 2^+_1$ is a clear outlier, with $R \simeq 0.11$. The observed discrepancy for the transitions involving the $4^+_2$ state may be related to mixing between the 
$4^+_2$ and $4^+_1$ states, the latter lying outside the 
IBM-CM model space~\cite{gavrielov2022zr}. For the decays of the $2^+_3$ state, the discrepancies are also large, but due to the larger experimental uncertainties they exceed 2$\sigma$ only for the $2^+_3 \to 2^+_1$ transition.}
The calculated spectroscopic quadrupole moments, reported in Table~\ref{tab:quadrupole_moments}, are {\color{black} smaller than the experimental result for the $2^+_1$, again, due to the fixed value of the effective $E2$ boson charge for the entire Zr chain, and agree for the $2^+_2$ state, which is almost twice the value}.

The results of the quadrupole sum rules analysis within the IBM-CM are reported in Table~\ref{tab:sum_rules} and shown in Fig.~\ref{fig:shapes}. The calculated $\langle Q^2 \rangle$ invariant for the $0_1^+$ state is slightly smaller than the experimental one, mainly because of the smaller calculated $B(E2;2^+_1 \to 0^+_1)$ value (see Table~\textcolor{black}{\ref{tab:transition_prob}}). However, both the experimental and theoretical $\langle Q^2 \rangle$ results provide a consistent picture of a quasi-spherical ground state exhibiting large quantum fluctuations in the overall deformation. The calculated and experimental $\langle Q^2 \rangle$ invariants for the $0_2^+$ state are in good agreement. For the $\langle \cos(3\delta) \rangle$ observable the IBM-CM yields the same sign (oblate) as the experimental results. 
\textcolor{black}{In both cases, the results translate to similar 
$\langle \delta \rangle$ values (see Fig.~\ref{fig:shapes}).}

To enable comparison with other calculations, it is possible to derive\footnote{$\langle \beta_2 \rangle=(\frac{3}{4\pi} Z R^2)^{-1}\cdot\sqrt{\langle Q^2 \rangle}$, where $R=1.2A^{1/3}$~[fm], and $\langle \gamma \rangle$ = $\langle \delta \rangle$~\cite{magdaeuroschool}. For $^{94}$Zr, $\langle \beta_2 \rangle \approx 0.35 \sqrt{\langle Q^2 \rangle}$.} from the model-independent quantities $\langle Q^2 \rangle$ and $\langle Q^3 \cos(3\delta) \rangle$ the usual deformation parameters $\beta_2$ and $\gamma$. The results show that the shapes obtained from the present experimental and IBM-CM results are also in \textcolor{black}{a reasonable} agreement with those predicted by the MCSM~\cite{togashi2016quantum,Leoni2024}: the T-plot analysis for $^{94}$Zr~\cite{Leoni2024} indicates that the ground state is spherical, and the $0^+_2$ state is oblate deformed with $\beta_2\lesssim 0.2$ (as compared to $\langle \beta_2 \rangle=\textcolor{black}{0.214(6)}$ resulting from the current experimental data) and a $\gamma$ value \textcolor{black}{in a 45$^{\circ}$-50$^{\circ}$ range, slightly larger than the experimental value}.

\section{Conclusions}
To summarize, this work reports the first experimental determination of shape parameters and their dispersions in a Zr isotope using quadrupole sum rules. The procedure, applied to the ground state and the first excited $0^+$ state in $^{94}$Zr, involved notably a measurement of spectroscopic quadrupole moments of the $2^+_{1,2}$ states, which represents a level of detail that is unique in the Zr nuclei. The rotational invariants obtained from the experimentally determined $E2$ matrix elements establish conclusively shape-coexisting structures in $^{94}$Zr, revealing a ground state with a shape that is not well defined, but tends towards sphericity \textcolor{black}{$(\langle Q^2\rangle=0.112(4)~e^2b^2$)}, and a $0^+_2$ state being more deformed \textcolor{black}{($\langle Q^2\rangle=0.371(21)~e^2b^2$) and triaxial-oblate shaped ($\langle \delta \rangle={34}(8)^{\circ}$)}. The invariants calculated within the IBM-CM model agree with the experimental findings and are also consistent with MCSM predictions for both states in question, supporting the QPT (IQPT) picture for a nucleus with $N<60$. Reaching a similar level of detail at $N=60$ and beyond will represent a challenge for new-generation radioactive-ion beam facilities.

\section*{Declaration of competing interest}
The authors declare that they have no known competing financial interests or personal relationships that could have appeared to influence the work reported in this paper.

\section*{Data availability}
Data will be made available on request.

\section*{Acknowledgements}
The authors thank the staff of the LNL Tandem-XTU accelerator for the excellent quality of the $^{94}$Zr beam, and M.~Loriggiola for producing the $^{208}$Pb target. The use of germanium detectors from GAMMAPOOL is acknowledged. This work was supported in part by the Natural Sciences and Engineering Research Council (Canada).

\bibliographystyle{elsarticle-num}
\bibliography{94Zr_GALILEO}

\end{document}